# Fano Resonances in a Photonic Crystal Covered with a Perforated Gold Film and its Application to Biosensing

**VASILY V. KLIMOV**[1,2,3,*]**, ANDREY A. PAVLOV**[1,2]**, ILYA V. TRESHIN**[1,2]**, AND ILYA V. ZABKOV**[1]

[1] *Dukhov Research Institute of Automatics (VNIIA), 22 Sushchevskaya Street, Moscow 127055, Russia*
[2] *P.N. Lebedev Physical Institute, Russian Academy of Sciences, 53 Leninsky Prospekt, Moscow 119991, Russia*
[3] *National Research Nuclear University MEPhI, 31 Kashirskoe Shosse, Moscow 115409, Russia*
*\* klimov@vniia.ru*

**Abstract**: Optical properties of the photonic crystal covered with a perforated metal film were investigated and the existence of the Fano-type resonances was shown. The Fano resonances originate from the interaction between the optical Tamm state and the waveguide modes of the photonic crystal. It manifests itself as a narrow dip in a broad peak in the transmission spectrum related to the optical Tamm state. The design of a sensor based on this Fano resonance that is sensitive to the change of the environment refractive index is suggested.

## 1. Introduction

Optical Tamm state, which can occur either on the boundary of two photonic crystals [1, 2] or on the boundary of a photonic crystal and a metal film [3–12], has recently attracted attention of researchers in nano-optics community. The optical Tamm state in a system with the photonic crystal covered with a metal film is especially interesting because it allows one to combine its complicated physics with the rich physics of plasmon surface waves. Since the first experimental demonstration [3], the optical Tamm state is an object of numerous researches. In particular, interaction of the Tamm state and a surface plasmon wave has been already shown [4, 5]. Besides, the scheme of a sensor which is sensitive to the change of refractive index based on the optical Tamm state has been offered [6, 7]. Recently, in a system consisting of a photonic crystal and a perforated metal film, effects of extraordinary transmission of light and huge asymmetry of transmission related to the optical Tamm state have been discovered [8, 9]. The optical Tamm state has been investigated in the presence of nonlinear effects, particularly as a lasing mode of the nanolaser [10], it has been used to enhance second harmonic generation [11] and to control spontaneous radiation of quantum emitters [12].

In this work, we investigate another effect associated with the optical Tamm state. The photonic crystal necessary to observe an optical Tamm resonance supports the propagation of a set of waveguide modes. In this paper, we show that it is possible to achieve an interaction between waveguide modes of the photonic crystal and the Tamm state with the help of a periodic lattice of slits in a metal film. We show that this interaction results in the Fano resonance, which is a typical sign of the interaction of high-quality dark and low-quality bright modes. In spite of the fact that the research of Fano's resonances in photonic and plasmonic structures is a hot topic now [13–17], we do not know any investigation devoted to the interaction of photonic crystal eigenmodes and optical Tamm state. The present article closes this gap.

The remaining part of the article is organized as follows. In section 2, eigenmodes of the dielectric photonic crystal covered with solid or perforated metal films are investigated. In section 3, transmission spectra through this optical system are presented, and the appearance of the Fano type resonances is shown. In section 4, we propose a new type of a sensor based on the sharp features of the Fano resonance observed in this system, which is sensitive to small variations of the adjacent analyte nanolayer refractive index. The geometry of the problem is shown in Fig. 1.

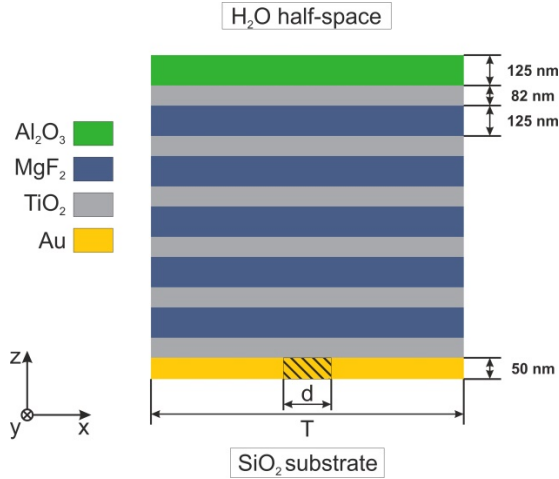

Fig. 1. The geometry of the problem. $N$ pairs of layers of $MgF_2$ and $TiO_2$ (the upper layer of $MgF_2$ is replaced by $Al_2O_3$) represent a photonic crystal. The gold film is either solid or perforated with a periodic array of slits of the width $d$ and the period $T$. The lower half-space is filled by silica, the upper half-space is filled by water.

## 2. Eigenmodes of the Dielectric Photonic Crystal Covered with a Solid or Perforated Metal Film

We study an optical device that consists of a gold film covered by a photonic crystal (see Fig. 1). The photonic crystal is composed of six pairs of layers of $MgF_2$ and $TiO_2$ with the topmost $MgF_2$ layer replaced by $Al_2O_3$. This structure resides on the $SiO_2$ substrate and is covered with water on top. Refractive indices of the $MgF_2$, $TiO_2$, $Al_2O_3$, $SiO_2$, and water are, respectively, $n_{MgF2} = 1.38$, $n_{TiO2} = 2.23$, $n_{Al2O3} = 1.63$, $n_{SiO2} = 1.45$ and $n_{H2O} = 1.33$. Refractive index of gold from Ref. 18 has been employed. The gold film is perforated with a periodic array of slits that are translational-invariant in $y$ direction. The slit width and pitch are $d$ and $T$, respectively. The slit volume is filled with $SiO_2$. Throughout the paper we will consider only $xz$ plane of incidence or $k_y = 0$, where $k_y$ is $y$-component of the wavevector $\mathbf{k}$ of the eigenmodes or the incident light.

In the previous works [8, 9], it has been shown that the optical Tamm state can be excited in this system at the vacuum wavelength of about 800 nm. This system also supports surface plasmon modes localized at the surfaces of the metal film and modes that are localized inside the photonic crystal — we will call them waveguide modes. The waveguide modes and the Tamm state can have either TM or TE polarization. Surface plasmon modes can only have TM polarization. Since we limit ourselves to the case of $k_y = 0$, for TM (TE) polarization the only non-zero component of the magnetic (electric) field is $H_y$ $(E_y)$. We will limit ourselves to considering TM-polarized modes only.

First, we consider the system without slits. In that case, the effective refractive indices of the eigemodes can be obtained by the Transfer Matrix method [19]. In this method, each mode is determined by the singularity (complex pole) in the transmission coefficient as a function of the complex frequency and the $k_x$ component of the wave vector. Figure 2 shows real parts of the effective refractive indices for Tamm state, surface plasmon, and waveguide mode in the system without slits.

In Fig. 2(a), we can identify three types of modes: surface plasmons (the red lines), waveguide modes of the photonic crystal (the green lines), and optical Tamm state (the blue line). The Tamm state has a cutoff wavelength of around 800 nm above which it does not exist. The waveguide modes also have cutoff wavelengths, but they are outside of the axes limits of Fig. 2. At the cutoff frequency, the $k_x$ component of the wavevector of the waveguide modes and the Tamm state is zero, which means that at these frequencies they are non-propagating in the $x$ direction and leak into the top and bottom half-spaces. The dashed lines in Fig. 2 correspond to the light lines in the upper and lower half-spaces — to the left of each of these lines modes are leaking in the corresponding half-space.

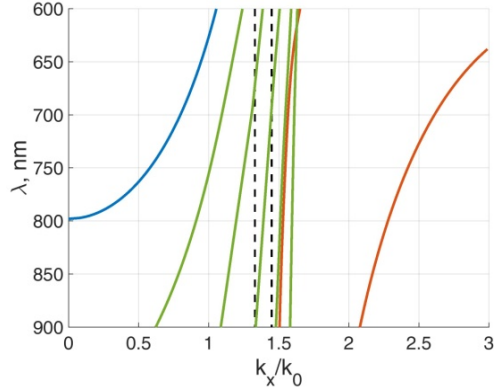

Fig. 2. Spectra of TM modes in the system shown in Fig. 1. The blue lines are the Tamm states, the red lines are the surface plasmons, the green lines are the waveguide modes of the photonic crystal, the dashed lines are the light lines in the upper medium (water, $n = 1.33$) and substrate (silica, $n = 1.45$).

Figure 3 shows the field distributions of the TM-polarized Tamm state (a) and one of the waveguide modes (b) along the $z$ direction. A characteristic feature of the Tamm state is high localization of the field near the interface of the metal and the photonic crystal (Fig. 3 (a)). The field of the waveguide mode is localized mainly on the opposite side of the photonic crystal (Fig. 3 (b)). Since the amplitudes of both modes are non-zero at the metal boundary, one may couple them by introducing an imperfection in the metal film, e.g. a slit. When two modes are coupled, one may expect to observe Fano-shaped resonance in the spectrum of such a system.

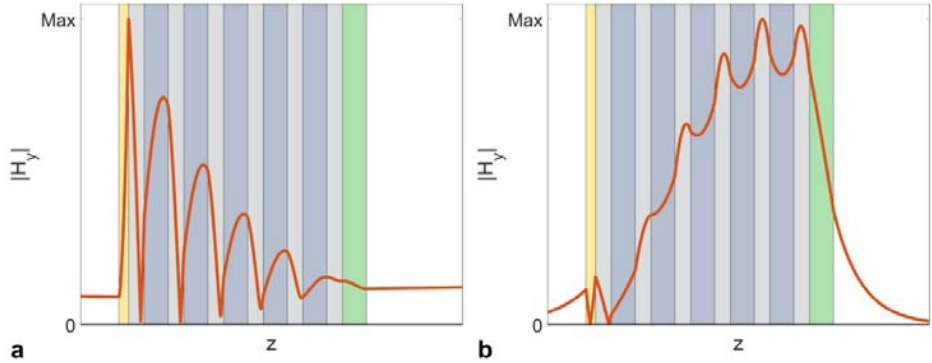

Fig. 3. Magnetic field distribution $\left(|H_y|\right)$ of TM polarized eigenmodes of the system: (a) Tamm state ($\lambda = 797.7$ nm, $k_x = 0$), (b) waveguide mode ($\lambda = 797$ nm, $k_x \approx 1.595 k_0$, rightmost waveguide mode at Fig. 2(a)).

In the presence of periodic array of slits, the relative spectral position of the Tamm state and the waveguide mode can be tuned by varying the pitch of the lattice. In the empty lattice approximation, the spectral position of the eigenmodes of the periodic array of slits can be evaluated from the following equation:

$$k_x^{\text{eig,Bz}} = k_x^{\text{eig}} + Gn, \quad (1)$$

where $k_x^{\text{eig,Bz}}$ is the $x$-component of the eigenmode's wavevector in the first Brillouin zone, $k_x^{\text{eig}}$ is the $x$-component of the eigenmode's wavevector in the absence of the array of slits, $G = 2\pi/T$ is the reciprocal lattice constant, and $n$ is an integer number ("diffraction order"). In order to couple the Tamm state and the waveguide mode of interest at the specified wavelength, the following condition must hold true ($n$ might be different for the Tamm state and the waveguide mode):

$$k_x^{\text{T,Bz}} = k_x^{\text{WG,Bz}}, \quad (2)$$

where $k_x^{\text{T,Bz}}$ and $k_x^{\text{WG,Bz}}$ are $k_x$ of the Tamm state and the waveguide mode in the first Brillouin zone, respectively. If we choose to intersect these modes at $k_x^{\text{T,Bz}} = k_x^{\text{T}} = 0$, i.e. at the normal incidence, the condition (2) reduces to:

$$k_x^{\text{WG}}\left(\lambda^{\text{T}}\right) = Gn, \quad (3)$$

where $\lambda^{\text{T}}$ is the cutoff wavelength of the optical Tamm state. In the case of the empty lattice approximation (when the slit width is infinitely small), this condition is exact. However, the increase in the slit width influences the spectral position of the modes making the condition (3) approximate.

In order to bring spectral features associated with TM-polarized Tamm state and waveguide mode from the Fig. 3(b) together at the wavelength that corresponds to $k_x^{\text{T}} = 0$, we have to choose the lattice constant to be $T = 500.2$ nm (according to equation (3)). Figure 4 shows the eigenmodes of such a system for $d = 5$ nm (white circles) together with the spectral position of the modes in the empty lattice approximation (1) (dashed lines) and with the absorption coefficient of the wave incident from $+z$ direction (background false-color image). The spectral position of the lattice eigenmodes for $d = 5$ nm and the absorption coefficient has been evaluated with the finite element method, implemented in COMSOL Multiphysics.

From the figure, one can conclude that while the spectral position of the waveguide mode coincides perfectly with the position predicted from the empty lattice approximation, the optical Tamm state position is shifted by approximately 3 nm from the predicted wavelength. This is due to the fact that for the Tamm state field localization is much higher in the metal film than for the waveguide mode and, hence, the waveguide mode is less perturbed by the presence of the slit than the Tamm state.

One may also notice from Fig. 4 that the positions of the eigenmodes are associated with the spectral features in the absorption coefficient. The Tamm state corresponds to the broad resonance, while the waveguide mode coincides with abrupt narrow resonances. This is related to the fact that quality factor of the Tamm state is about 100 times lower than that of the waveguide mode. As a quality factor for this kind of eigenmodes, we understand the following: $Q_{\text{eig}} = \omega'_{\text{eig}}/2\omega''_{\text{eig}}$, where $\omega_{\text{eig}} = \omega'_{\text{eig}} + i\omega''_{\text{eig}}$ is the complex eigenfrequency of the eigenmode. In the next section, we will discuss spectral features that arise from the interaction between the Tamm state and waveguide mode in such system more thoroughly.

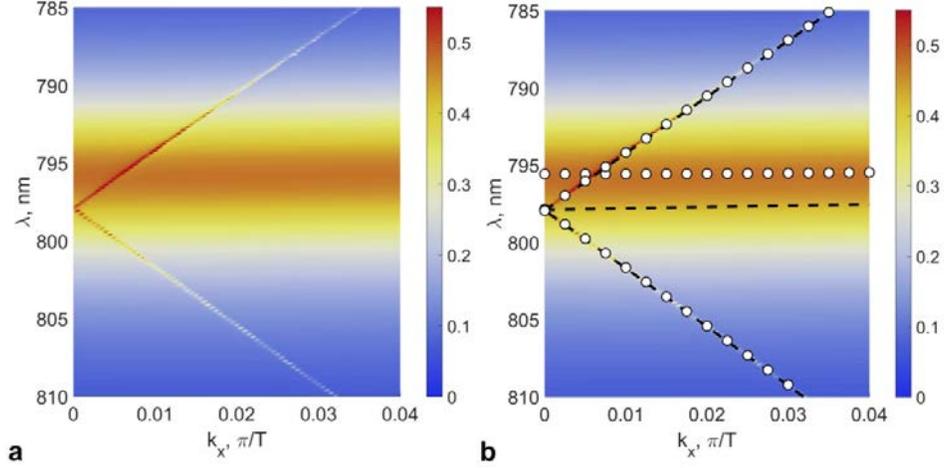

Fig. 4. (a) Absorption coefficient of the wave incident on the structure with $N = 6$, $T = 500.2\,\text{nm}$, and $d = 5\,\text{nm}$ versus the vacuum wavelength and $k_x$ of the incident wave (TM polarization). (b) Background: the same as in the subfigure (a), dashed lines: spectra of the optical Tamm state (Fig. 3(a)) and the waveguide mode (Fig. 3(b)) in the first Brillouin zone in the empty lattice approximation. White circles: spectra of the optical Tamm state (Fig. 3(a)) and the waveguide mode (Fig. 3(b)) in the presence of the slit array.

## 3. Transmission Spectra of a Dielectric Photonic Crystal Covered with a Metal Film with a Periodic Array of Slits

Transmission coefficients for the normal TM-polarized light incident on the structure with slits ($d = 5\,\text{nm}$) are shown in Fig. 5(a) for different lattice periods. The asymmetric shape of the transmission curve is a clear indication of the Fano resonance. From the figure one can see that the shape of the Fano resonance changes while high-quality waveguide mode is moving across wide Tamm state resonance.

For qualitative understanding of the transmission spectrum through our system, we have use the model of two coupled classical oscillators one of which is driven by the external force $f(t) \sim e^{i\omega t}$:

$$\begin{cases} \ddot{x}_1 + 2\gamma_1 \dot{x}_1 + \omega_1^2 x_1 + \kappa x_2 + \nu \dot{x}_2 = f(t), \\ \ddot{x}_2 + 2\gamma_2 \dot{x}_2 + \omega_2^2 x_2 + \kappa x_1 + \nu \dot{x}_1 = 0. \end{cases} \quad (4)$$

The first oscillator ($x_1$) corresponds to the low-Q Tamm state that is excited directly. The second oscillator ($x_2$) corresponds to the high-quality waveguide mode which is excited through the interaction with the Tamm state. Eigenfrequencies $\omega_1$ and $\omega_2$ and decay rates $\gamma_1$ and $\gamma_2$ are approximately equal to the eigenfrequencies and decay rates of the optical Tamm state and waveguide mode, respectively. By varying the coupling constants $\kappa$ and $\nu$, we fit the amplitude of the first oscillator to the transmission spectrum of the structure. Amplitudes of the first oscillator which are related to the transmission for different eigenfrequencies of the second oscillator are shown in Fig. 5(b). From comparison of Fig. 5(a) and 5(b), one can see fine agreement between the model and the results of numerical calculations. It indicates that our understanding of this phenomenon is correct.

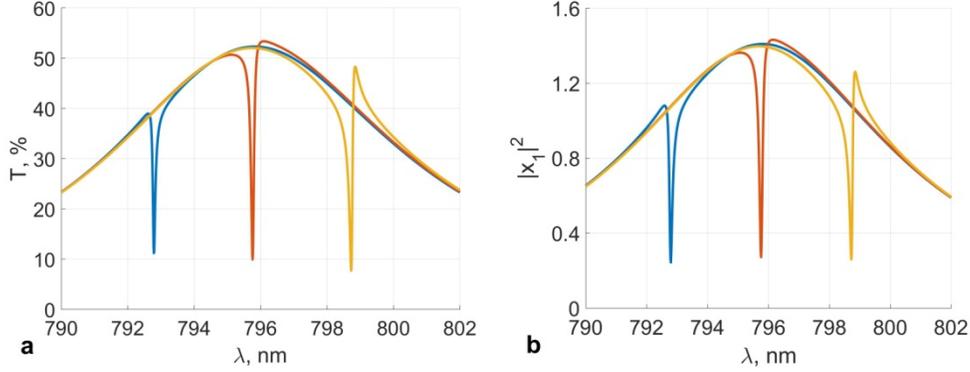

Fig. 5. (a) Fano resonances in transmission coefficient through the system for three lattice periods: 496.8 nm (blue), 498.8 nm (red), and 500.8 nm (yellow). (b) Coupled oscillator model fit of the transmission curves. The only parameter that is different for three fits is eigenfrequency of the high-Q oscillator.

From Fig. 5, one can see that while eigenfrequency of the waveguide mode is being tuned across the wide Tamm state by varying the lattice pitch, the shape of the Fano resonance is changing and even when resonance frequencies of the Tamm state and waveguide mode coincide, the resonance has an asymmetric shape. This is due to the non-zero coupling constant $v$. In principle, engineering this coupling coefficient can lead to very sharp Fano resonances with extremely narrow spectral features. Even greater control over the shape of the resonance can be obtained if other modes are introduced into the system and coupled to the Tamm state.

Fano resonances from Fig. 5 can be used to detect small variations in the refractive index of the surrounding media. It can serve as an alternative to the popular surface plasmon based sensors and has some advantages over them. One of them is extremely narrow dips that are attainable in such system. Another is the ability to work at normal incidence and measure transmitted, not reflected light. In the following section we will discuss the properties and performance of such Fano resonance-based sensor.

## 4. The Refractive Index Sensor Based on a Dielectric Photonic Crystal Covered with a Metal Film with a Periodic Array of Nanoslits

Utilizing optical nanostructures that support highly localized plasmonic and photonic modes is a popular approach in biological and chemical sensors design [20]. We believe that our structure can have a competitive performance when compared to other analogous devices. Sharp features of the Fano resonance have already been applied to detect a small variation of the refractive index [21–23]. Our system has a substantially pronounced Fano-type resonance in comparison with those works because of a high quality of the waveguide modes of the photonic crystal. That is why one can expect that our system can result in a more sensitive detector of the refractive index.

To estimate the sensitivity of our device, we will define it in relation to the refractive index $n_a$ of a thin layer of analyte (layer thickness is 50 nm, see Fig. 6). Otherwise, its configuration is similar to the configuration studied in the previous sections. Let us note that unlike the case of the analyte occupying the half-space (see, for example [6]), this formulation of the problem is closer to actual sensing experiments (see, for example [21]). Thickness on the order of tens of nanometers is dictated by the size of the molecules of interest and the functionalization scheme. For example, a single layer of low density lipoproteins with functionalization layers will have the thickness of 30–50 nm.

The most common sensitivity criterion is the shift of the resonance wavelength per refractive index change: $S_\lambda = \Delta\lambda/\Delta n_a$, where $\Delta\lambda$ is the shift in the resonance wavelength when the refractive index of analyte changes by $\Delta n_a$. However, in order to exploit the sharp shape of the Fano resonance in our system, it is reasonable to use the criterion that is associated with the transmission intensity change at the fixed wavelength. In this case one of the two following figures of merit can be considered:

$$S_T = \frac{\Delta T/T_0}{\Delta n_a}, \qquad (5)$$

$$\tilde{S}_T = \frac{\Delta T}{\Delta n_a}, \qquad (6)$$

where $T_0$ stands for the transmission coefficient without analyte, while $\Delta T$ is the change in transmission coefficient when refractive index of 50 nm thickness analyte changes by $\Delta n_a$. Definition (5) has been used in the Ref. 20, which we will use when evaluating performance of our sensor is compared to the state of the art sensors.

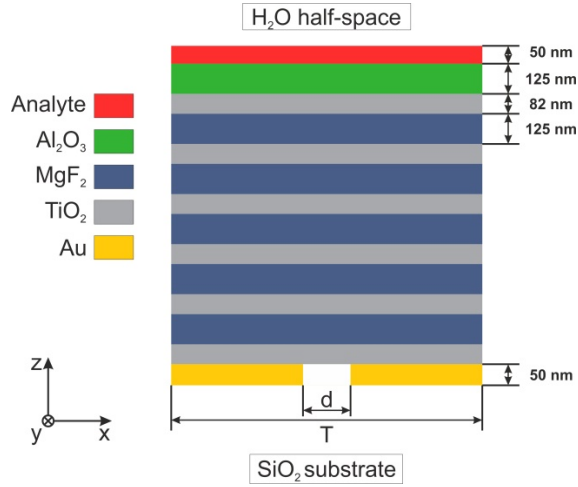

Fig. 6. The configuration of a sensor similar to one of Fig. 1, but with a thin layer of an analyte on the top of the photonic crystal. The slit width is $d = 5$ nm, the lattice constant is $T = 498.8$ nm.

The resonance wavelength dependence for the system in Fig. 6 on the analyte refractive index is presented in Fig. 7(a). From the figure, it is clear that in the studied range of refractive indices the resonance shifts linearly with the refractive index. From the least-squares approximation, the wavelength sensitivity of this device can be estimated as $S_\lambda \approx 10.75 \, \text{nm/RIU}$, which is relatively low value compared to analogous devises. Sensitivity of state-of-the art plasmonic sensors (Ref. 24, 25) is about $30000 \, \text{nm/RIU}$. However, the detection limit of such devices is related to the width of the resonance, and the plasmonic resonance usually has the width of about 50–100 nm. Consequently, one can introduce figure of merit (FOM) which is the sensitivity divided by the full width at half maximum of the sensitive resonance. For our device, the FOM $\approx 233 \, \text{RIU}^{-1}$, while for state-of-the-art sensors it is 330 RIU$^{-1}$ [24] and 590 RIU$^{-1}$ [25]. Sensitivity and the FOM values of our device can be even better if we consider the analyte half-space instead of 50 nm layer.

Dependence of the transmission intensity at fixed wavelength on the refractive index is presented in Fig. 7(b). As one can see, this dependence is nonlinear and saturates for large values of $\Delta n_a$. However, for the values of $\Delta n_a < 0.01$ we can evaluate linear approximation which gives us sensitivity $S_T \approx 33700\,\%/\text{RIU}$ and $\tilde{S}_T \approx 3340\,\%/\text{RIU}$. For the state-of-the-art sensor that operates in the intensity sensitivity regime this values are $S_T = 48117\,\%/\text{RIU}$ [21] and $\tilde{S}_T \approx 260\,\%/\text{RIU}$ (our estimation based on the information from the Ref. 21).

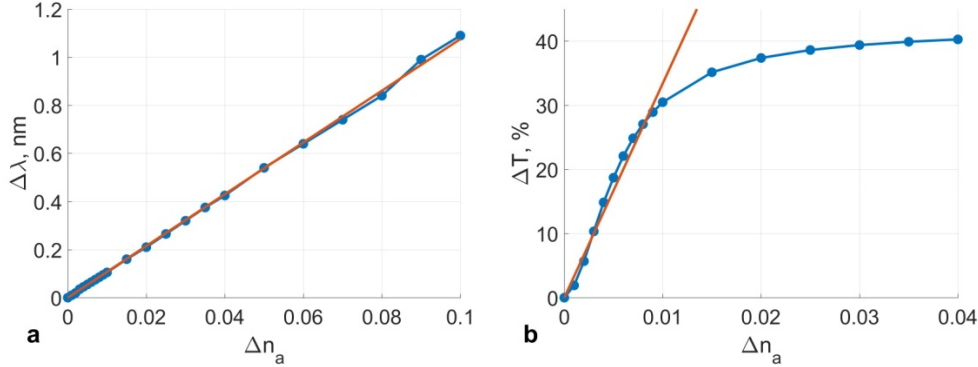

Fig. 7. (a) The shift of the Fano resonance due to the change of the analyte refractive index (blue dots) and its linear approximation (red line). The linear approximation gives the wavelength sensitivity of 10.75 nm/RIU. (b) The change in transmission signal at a fixed wavelength versus the change in the analyte refractive index (blue dots) and its linear approximation before the saturation. The linear approximation gives the transmission sensitivity of 33700%/RIU.

Let us note that by varying number of layers, their thicknesses, and their material properties, one can finely tune quality factors and field localization of the waveguide modes in the system, and the sensitivity of the sensor based on such system will change accordingly. We have run optimization procedure for the wavelength sensitivity of the waveguide mode with layer thicknesses and their number as a variable that is we have maximized the function $S_\lambda(N, h_{\text{Al2O3}}, h_{\text{MgF2}}, h_{\text{TiO2}})$. In order to speed up the process, we have performed calculations of the wavelength sensitivity in the absence of the holes (in the empty lattice approximation) via T-matrix method, implying that sensitivity won't change significantly in their presence. In the Fig. 8, the wavelength sensitivities and quality factors for the waveguide eigenmodes found during the optimization process are depicted, each point corresponds to a different device configuration. The higher are the wavelength sensitivity $S_\lambda$ and the quality-factor $Q$ of the waveguide mode, the higher is the transmission sensitivity $S_T$ (or $\tilde{S}_T$). Therefore, one can achieve sensitivity in transmission signal as high as $10^6\,\%/\text{RIU}$ according to our calculations for some of the points from Fig. 8. However, this comes at a price of extremely narrow resonance widths with Q-factor on the order of $10^6$, which are extremely hard to attain in practice due to fabrication limitations (see, however, Ref. 26 for the example of high-Q Bragg cavity). Thus, practical performance of the sensor which is based on the proposed configuration will be limited by the quality of the fabricated structures since in theory one can achieve almost arbitrary values for sensitivity in the range from hundreds to millions percent per RIU. As one can see from the Fig. 8, a large portion of the overall configurations have the wavelength sensitivity around 17 nm/RIU and quality factor of about $3\times10^4$. This tells us that our device's performance might be very robust in this region.

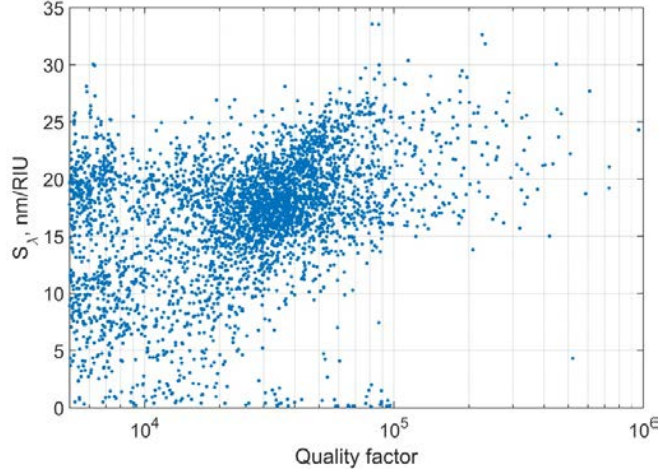

Fig. 8. Illustration of the optimization process. Each point on the graph represents a waveguide mode quality factor and sensitivity for certain configuration of the sensing device with variable $MgF_2$, $TiO_2$ and $Al_2O_3$ thicknesses and layer numbers. Q-factors and sensitivities of the eigenmodes have been found in the absence of the holes via the T-matrix method.

Let us note that high sensitivity can be also achieved in a photonic crystal without metal [27, 28]. However, in this approach the Kretschmann configuration of attenuated total internal reflection should be used to excite the waveguide modes. This geometry can be hardly used for multimodal sensors (lab on the chip). Instead, we suggest to use a periodic lattice of slits in a metal film to provide the interaction between the Tamm state and waveguide modes. A significant advantage of such a configuration of the sensor is a possibility to work in transmission regime with excitation of the sensor by a wave with normal incidence. This allows one to place multiple sensors with 200 μm footprint each on one chip. An artistic vision of our sensor is shown on Fig. 9.

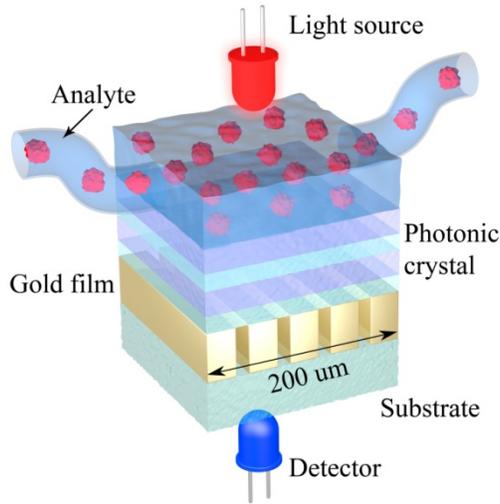

Fig. 9. Artist view of the refractive index sensor working in transmission mode.

## 5. Conclusion

Eigenmodes of a photonic crystal covered with a solid and perforated metal film have been found and investigated. Coupling of the optical Tamm state with waveguide modes via a peri-odic array of slits has been demonstrated. In particular, this coupling effect manifests itself as a narrow Fano-type resonance in transmission spectrum. Configuration of a sensor sensitive to the variations of the refractive index of a 50 nm layer (which corresponds to, e.g., a single layer of low density lipoproteins) of analyte based on this phenomenon has been proposed. Working with TM-polarized normally incident light and detecting transmitted rather than reflected signal are the advantages of this sensor. A narrow shape of the Fano resonance allows one to achieve the sensitivity of $33700\,\%/\text{RIU}$ for transmission intensity. This sensitivity can be significantly improved from the theoretical point of view by varying the thicknesses of the layers and their material properties. The only limitation of our sensor sensitivity is due to technological issues of photonic crystal production. For the state-of-the-art photonic crystal the sensitivity can be as high as $10^5\,\%/\text{RIU}$.

## 6. Funding

The research has been supported by the Advanced Research Foundation (Contract No. 7/004/2013-2018). The authors are also grateful to the Russian Foundation for Basic Research (Grants No. 14-02-00290 and No. 15-52-52006) for financial support of this work.

**ACKNOWLEDGEMENTS**

Authors would like to thank A. V. Baryshev and A. M. Merzlikin for the valuable discussions and comments.